# Correlative Symmetric Index: An alternative mathematical evaluation for beam profile symmetry

Author: Dalton H Bermudez[1], Wesley Culberson[1]
[1] Medical Physics Department, University of Wisconsin-Madison

Corresponding:
Dalton H Bermudez
dbermudez@wisc.edu

**Abstract**

**Background:** Current mathematical quantification methods for beam symmetry are highly sensitive to noise, particularly in beam profiles with significant variations.

**Purpose:** This study evaluates the accuracy of standard radiotherapy beam symmetry metrics and compares them to a proposed cross-correlation-based metric called the Correlative Symmetric Index (CSI) and the Structural Similarity Index (SSIM). We aim to demonstrate that CSI is less susceptible to noise than the traditional methods.

**Methods:** Simulated non-symmetric beam profiles with similar left and right areas were analyzed using both the standard and proposed symmetry metrics. To test the robustness of each method to noise, a noisy non-symmetric beam profile was generated. The measured beam profiles at various depths in the water tank were used to compare the performance of each symmetry metric under realistic clinical conditions.

**Results:** In the noisy and non-symmetric cases, the CSI and SSIM values were 0.387 and 0.5401, respectively. Traditional metrics such as the Point Difference Quotient (PDQ) and area-based symmetry yield values of 0.312 and 0.400, respectively. The percentage changes between the non-symmetric (no noise) and non-symmetric (with noise) cases for PDQ, area-based symmetry, SSIM, and CSI were 11.1%, 5.6%, 16.25%, and 1.43%, respectively. There was a statistically significant difference in the percentage chance for various model noise profiles between the proposed CSI metric and the standard metric. For clinically measured symmetric profiles, all metrics produced values above 0.9.

**Conclusion:** The Correlative Symmetric Index (CSI) demonstrates greater robustness to noise than both pointwise and area-based symmetry methods. CSI provides a reliable method for quantifying beam symmetry, particularly under noisy conditions.

Funding Statement: This work was supported by the National GEM Consortium Fellowship and the SciMed GRS provided by the Graduate school, part of the office of Vice Chancellor for research and Graduate Education.

Ethical Compliance: All the data was from Simulated datasets and beam profile acquired in a water tank. No ethical concern.

Conflict of Interest declaration: The authors declare that they have no affiliations with or involvement in any organization or entity with any financial interest in the subject matter or materials discussed in this manuscript.

## Introduction

Accurate quantification of beam symmetry is vital to ensure the quality of radiotherapy. Traditional beam symmetry metrics, such as the Point Difference Quotient (PDQ) and area-based symmetry, have long been used to characterize symmetry, but are often sensitive to noise and may not fully reflect the visual similarity of beam profiles. A more perceptually relevant and noise-resilient method may enhance the detection of clinically significant asymmetries.

This study introduces a Correlative Symmetric Index (CSI), which is a mathematical metric based on the cross-correlation of the left and right halves of a beam profile. In addition, we evaluated the Structural Similarity Index (SSIM), a perceptually motivated measure from the image processing domain, to assess its performance in the context of beam symmetry evaluation. Our hypothesis is that both CSI and SSIM will offer improved robustness to noise and greater fidelity to perceptual symmetry than conventional methods.

## Methods:

### *Standard approaches for beam symmetry assessment*

Beam symmetry is described as the area of the radiation distribution within the crosspatch of the beam profile. It allows a valuable understanding of the uniformity of the radiation field and detects asymmetries or nonuniformities. The procedure generally starts by obtaining a 2D profile of the intensity of the beam and evaluating an area within the profile. The beam profile can be acquired with a radiation detector, such as an array or detector scanner, by detecting the radiation intensity at points distributed in the cross-section of the beam. [1] The beam profile data obtained may be used in a raw state or, alternatively, may be smoothed using smoothing functions to reduce overall beam intensity variations, allowing corresponding points to be directly compared. One way to quantify the symmetry of a beam profile is PDQ calculation (Eq(s) 1 and 2).

$$Symmetry = 100 * max \left\{\frac{D_{i,left}}{D_{i,right}}, \frac{D_{i,right}}{D_{i,left}}\right\} \quad i=0,1,k-1, \qquad \text{(Eq 1)}$$

$$PDQ = max \left((D_{i,right} - D_{i,left})/max\,(D_{i,left,}, D_{i,right})\right) \qquad \text{(Eq 2)}$$

The point difference quotient (PDQ) algorithm assesses beam symmetry by comparing the dose or intensity levels at corresponding points on each side of the central axis in the beam profile. This type of investigation allows for local analyses of inhomogeneity and is sensitive to dose differences at specific points. However, its applicability in real practice has a restriction in that it may be very sensitive to noise in measurement data, and even small fluctuations will result in dramatic variations in PDQ values. Other factors, including the calibration error and positioning error of the detector, can cause more unreliability in the measurement of the PDQ. Although it is sensitive for the identification of localized deviations, it may not reflect global deviations in beam symmetry. Its performance is also sensitive to the spatial resolution of the collected data because sparsely sampled profiles cannot accurately reflect the overall symmetry. Additionally, detecting exact matching points across the central axis may be difficult for complex or irregularly shaped beams. In any case, most of these 3D measurement systems provide automated centering functionalities that facilitate post-acquisition profile alignment, so this problem can be minimized. [2]

Another approach for characterizing beam symmetry is the similarity of the left and right area quotients (Eq. 3) [3].

$$S = 100 * \frac{(area_{left} - area_{right})}{(area_{left} + area_{right})} \qquad \text{(Eq 3)}$$

The area quotient method evaluates beam symmetry by comparing the areas on either side of the central axis, defined at half the maximum intensity point. In a symmetric beam or a uniform dose distribution, the areas on both sides are equal. Unlike the Point Difference Quotient (PDQ) approach, this method offers a global assessment of beam asymmetry, but may overlook localized variations or hotspots within the profile. Its accuracy is influenced by the spatial resolution of the radiation detector, and detectors with larger sizes and lower resolutions may fail to capture fine asymmetries. Additional factors such as detector positioning, setup conditions, and calibration precision also affect the reliability of this method. Furthermore, the area quotient is not sensitive to symmetry variations along different angles or orientations, particularly in the case of nonuniform beam profiles.

***Correlative Symmetric Index***

The proposed approach characterizes beam symmetry using cross-correlation [5], dubbed the Correlative Symmetric Index (CSI). Eq(s) 4 describes the mathematical methodology.

where the value of CSI ranges from 0 to 1, with one being the perfect linear relationship between the two halves of the profiles and 0 indicating no linear relationship between the two halves of the beam profile. Here, $D_{rigth}$ refers to the array of intensity values measured in the right half of the beam profile (see range 0–6 in Figure 1). $D_{left,mirror}$ represents the other half of the beam profile, mirrored across the y-axis (or flipped horizontally) (as in range -4 through 0 in Figure 1) in reverse order, and $n$ is the total number of data points collected. To find the CSI, $D_{right}$ is multiplied by each of its points while it is shifted by m points (lagged) to the corresponding points in $D_{left,mirror}$ and then all multiplied points are added for each corresponding m lag value. To determine the most significant values of similarity between the two distributions, a lag of zero is the most valuable CSI value. The simulated non-symmetric distributions of a biased CSI were used to give a more representative value for the non-symmetric distribution (Eq. 5), and the acquired clinical symmetric distributions used as well as biased CSI values were calculated (Eq. 5). However, a slightly different normalized CSI can also be incorporated, as expressed in Eq(s) 6. where the biased CSI is calculated by dividing the CSI in Eq(s) 4 by the number of points in the distribution, while the normalized CSI is calculated by dividing the CSI in Eq(s) 4 by the square root of autocorrelations of $D_{left,mirrior}$ and $D_{left,mirrior}$ for lag of zero. We incorporated an additional measurement of structural similarity between the two halves of the profiles to serve as a supplement form of quantification to that of the correlative symmetric index (Eq 7). where $\mu_{D_{rigth}}$ is the average of distribution $D_{right}$, $\mu_{left,mirror}$ is the average of distribution $D_{left,mirrior}$, $\sigma^2_{D_{leftmirror}}$ is the variance of $D_{left,mirrior}$, $\sigma^2_{D_{rigth}}$ is the variance of $D_{right}$, $\sigma_{D_{rigth}D_{left,mirror}}$ is the covariance of $D_{left,mirrior}$ and $D_{right}$ and c1 and c2 are two variables to stabilize the division with a weak denominator. The values of constants c1 and c2 were obtained as shown in Eq(s) 8, where L is the specific dynamic range of the intensity profiles. The structural similarity index measure (SSIM) is a measurement of the similarity between two halves of a beam profile distribution. The SSIM values ranged between 0, suggesting no structural similarity, and 1, indicating perfect structural similarity. These mathematical Eq(s) were implemented using MATLAB xcorr and SSIM functions.

$$\text{CSI} = (D_{left,mirror} * D_{rigth})[n] = \sum_{-\infty}^{\infty} \underline{D_{left,mirror}[m]} D_{rigth}[m+n] \qquad \text{(Eq 4)}$$

$$CSI_{biased} = \frac{1}{N}\text{CSI} \qquad \text{(Eq 5)}$$

$$CSI_{normilized} = \frac{1}{\sqrt{CSI_{D_{left,mirror}D_{left,mirror}}(0)\,CSI_{D_{Rigth}D_{Rigth}}(0)}}\text{CSI.} \qquad \text{(Eq 6)}$$

$$\text{SSIM}(D_{rigth}, D_{left,mirrior}) = \frac{\left(2\mu_{D_{rigth}}\mu_{D_{left,mirror}}+c_1\right)(2\sigma_{D_{rigth}D_{left,mirror}}+c_2)}{(\mu^2_{D_{rigth}}+\mu^2_{D_{left,mirror}}+c_1)(\sigma^2_{D_{rigth}}+\sigma^2_{D_{leftmirror}}+c_2)} \qquad \text{(Eq 7)}$$

$$C_1 = (0.01L)^2$$

(Eq8)

$$C_2 = (0.03L)^2$$

Extrapolating the beam symmetry from a covariance-based approach allows characterization that is sensitive to both local and global deviations, overcoming the limitations of previously accepted standard approaches. Maintaining a pointwise analysis, as in the PDQ approach to beam symmetry characterization, retains the ability of the method to detect smaller-scale asymmetries between two halves of the beam profile. Similarly, performing an area-wise comparison between the two distribution halves via mirroring allows the CSI method to identify larger-scale variations in the net area under the curve for both, akin to the Similarity of Left and Right Area approaches. Additionally, the CSI metric was found to be more robust under noisy data collection compared to the standard methods, as depicted in the percentage difference calculations between this method and other more well-known methods between the obtained metrics for the non-symmetric (no noise) and non-symmetric (with noise) models in our results.

***Model of non-symmetric beam***

Uniform and non-uniform beams were simulated to demonstrate, with fine control, the parameters that influence both the standard and proposed symmetry metrics. This study predominantly focused on the effects of symmetry deviations and noise to evaluate the effectiveness and robustness of the existing and proposed beam symmetry characterization techniques.

The asymmetric beam profiles (Figure 1) were simulated using a superposition of phase-shifted sine and cosine functions, as described in Eq(s) 9-11. Asymmetry was introduced via a phase shift, moving the center of the otherwise symmetric distribution off the axis of symmetry. One can notice the asymmetry of this model beam by comparing the area specified by the number of grided squares underneath the curve towards the left of the zero-position axis with the number of ground squares underneath the curve for anything towards the right of the zero axis.

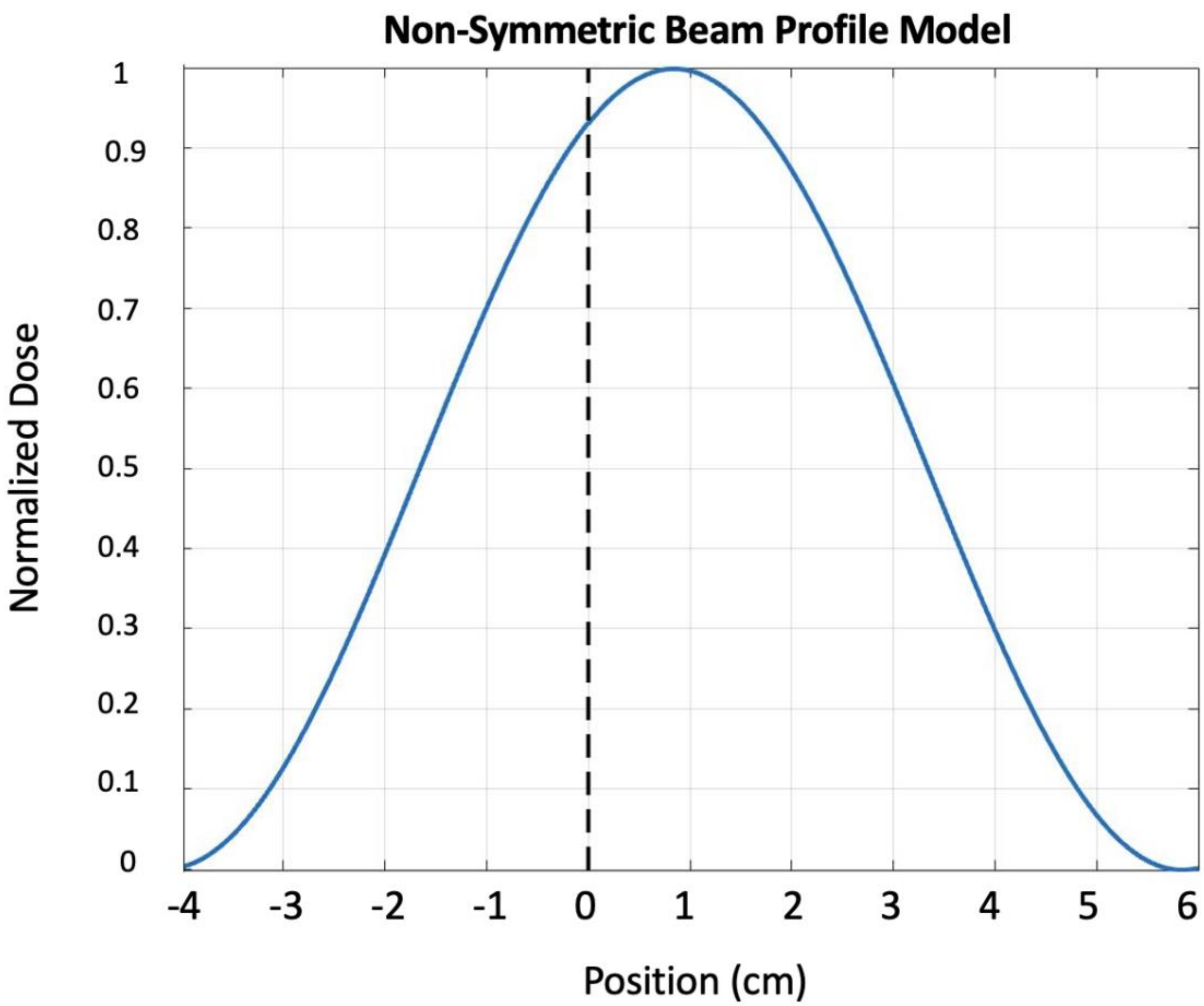

**Figure 1:** Simulated beam profiles were modelled as a superposition of sinusoids with varying amplitudes and phase delays (see Eq(s) 9). Asymmetry was introduced by shifting the distribution center, simulating a potential error in practical measurements of beam symmetry.

$$Beam\ Profile = A1 * sin\ sin\ (k * x + \phi)\ + A2 * cos\ (k * x) \qquad \text{(Eq 9)}$$
$$Beam\ Profile = Beam\ Profile - Min(Beam\ Profile) \qquad \text{(Eq 10)}$$
$$Beam\ Profile = Beam\ Profile./Max(Beam\ Profile) \qquad \text{(Eq 11)}$$

In Eq(s) 5, *A1* and *A2* are constants that dictate the amplitudes of superimposed sinusoidal waves. The variable $k$ corresponds to the spatial frequency of each wave. The asymmetry-introduced phase shift imposed on the sine wave contribution is represented by the variable $\phi$. For these beam profile simulations, the parameters were A1=1, A2=0.5, k = $2\pi/10$, and $\phi = \pi/4$. Then, the model was shifted so that it started at zero (Eq 10) and normalized by the maximum value of the modelled beam (Eq 11) by performing an element-wise division between the maximum beam profile value and all corresponding points in the beam profile distribution.

***Model of realistic non-symmetric beam***

The purpose of these simulations is to provide a more accurate depiction of the non-symmetric beam profile based on the effects that can be found in an actual treatment setting [6]. Different beam symmetric metrics, both traditional and proposed, were examined in this model of a more non-symmetric beam that could be encountered in a clinical setting [7]. This enables us to depict the effectiveness of our proposed method compared with traditional methods for quantifying the non-symmetric nature of a beam found in a clinical setting.

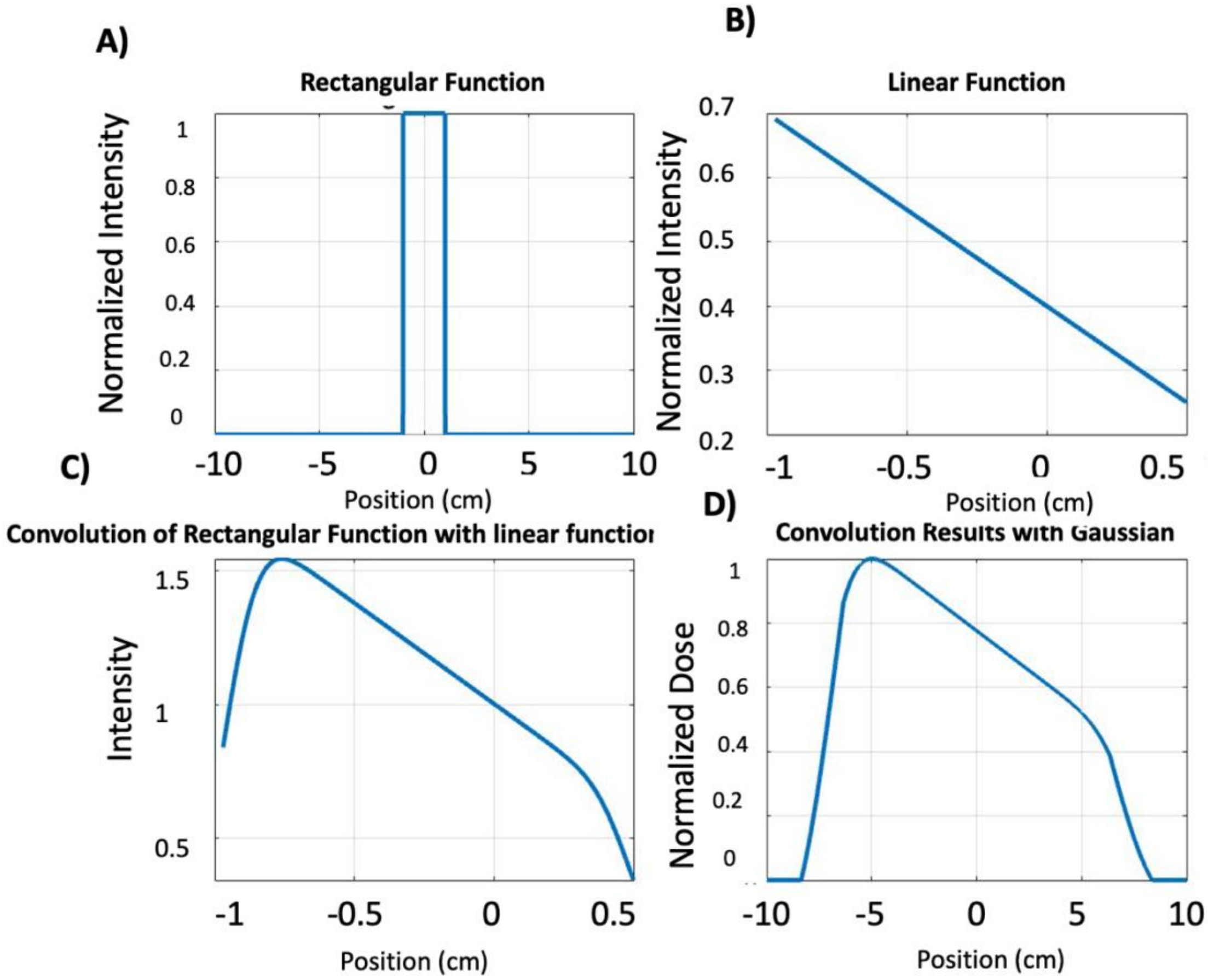

**Figure 2: A)** Rectangular function used for simulation of beam Wedge effects **B)** Linear function used for simulation of beam Wedge effects. **C)** Convolution result between Gaussian function and linear function. **D)** Convolution result of the correlation between Gaussian/Linear function (C) and the rectangular function (A).

The wedge-based non-symmetric beam profile was modelled as a convolution of a Gaussian function with a linear function (See Eq(s) 12) (Figure 2C). The result of the previous convolution was then convoluted using a rectangular function (Figure 2D) (Eq(s) 13).

$$\left(f(x) * l(x)\right) = \sum_{-\infty}^{\infty} e^{\left(\frac{-\tau^2}{2\sigma^2}\right)}(-0.3(x-\tau)+0.4)) \quad \text{(Eq 12)}$$

$$g(x) * \left(f(x) * l(x)\right) = \sum_{-\infty}^{\infty} \Pi(\tau) \sum_{-\infty}^{\infty} e^{\left(\frac{-\tau^2}{2\sigma^2}\right)}(-0.3(x-\tau)+0.4) \quad \text{(Eq 13)}$$

$$g(x) * \left(f(x) * l(x)\right)./max(g(x) * \left(f(x) * l(x)\right)) \quad \text{(Eq 14)}$$

The defined position vector for both the rectangular (Figure 2A) and Gaussian functions ranged from -10 to 10 in intervals of 0.01. The defined position vector for the linear function (Figure 2B) was in the range of -0.97 to 0.5 with a sampling interval of 0.001. The $\sigma = 1$ for the Gaussian function. The model was then normalized by the maximum value of the modelled beam (Eq 14) by performing an element-wise division between the maximum value of the distribution and all other points in the beam profile distribution.

***Measure beam profiles***

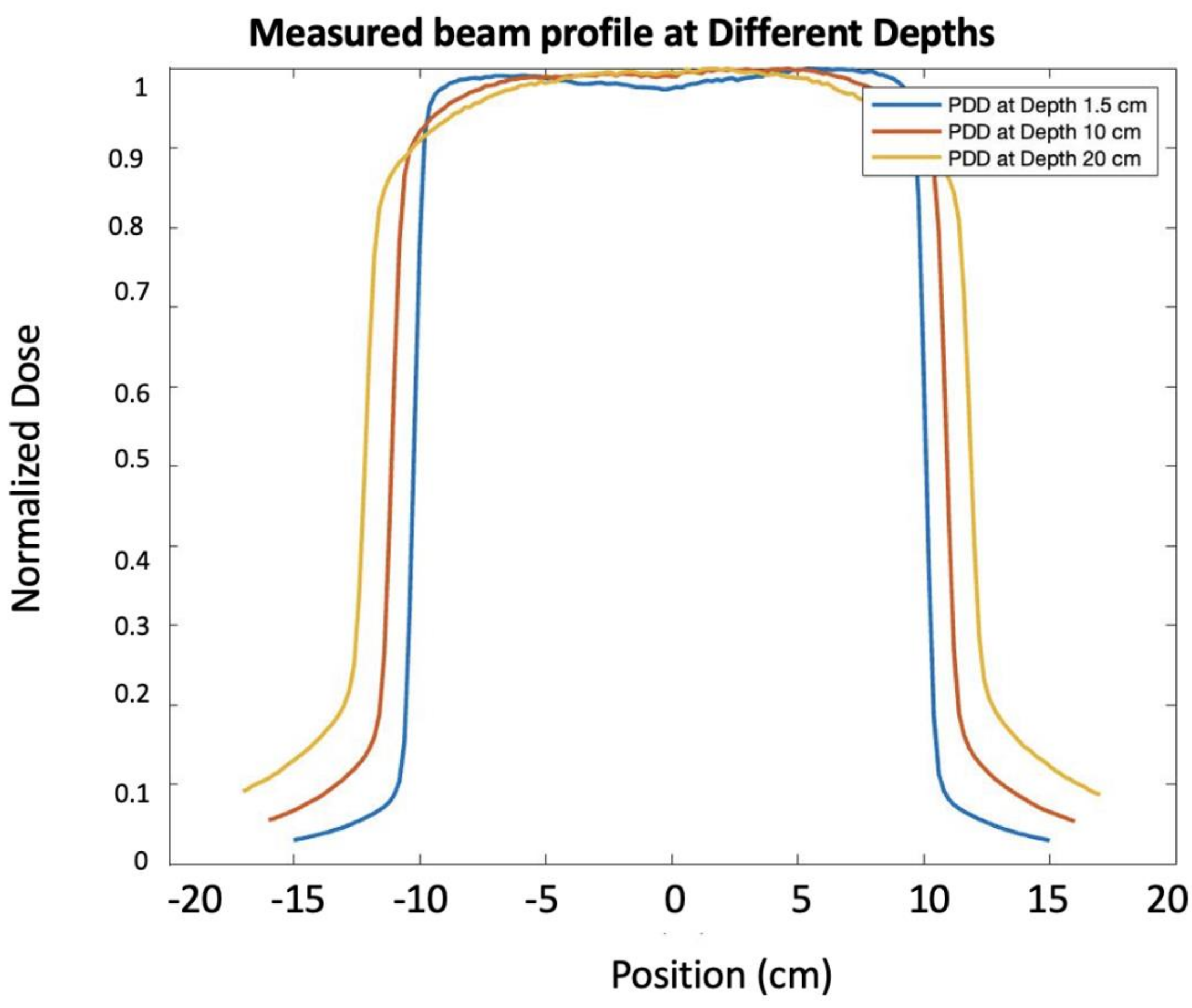

**Figure 3:** Measured beam profiles at depths of 1.5 cm, 10cm, and 20 cm

To validate the accuracy and robustness of the proposed CSI metric, analyses were performed on a series of simulated beam profiles to highlight the specific advantages of this approach over the other methods. Additional analyses were performed using beams generated with a Varian TrueBeam linear accelerator and profiles recorded with a scanning ion chamber, the 0.125cc

Exradin A28 (Standard Imaging, Middleton, WI), according to the protocols outlined in AAPM TG-106, to demonstrate the applicability of the proposed method in practical usage.

To assess the accuracy of the standard metrics and the proposed Correlative Symmetric Index method under realistic conditions, each was used to evaluate a real beam profile that was obtained in a water tank, following the protocols of TG-106 at depths of 1.5 cm, 10 cm, and 20 cm at 6MV 20x20 cm$^2$ at 100 SSD (**Figure 3)**.

**Results**

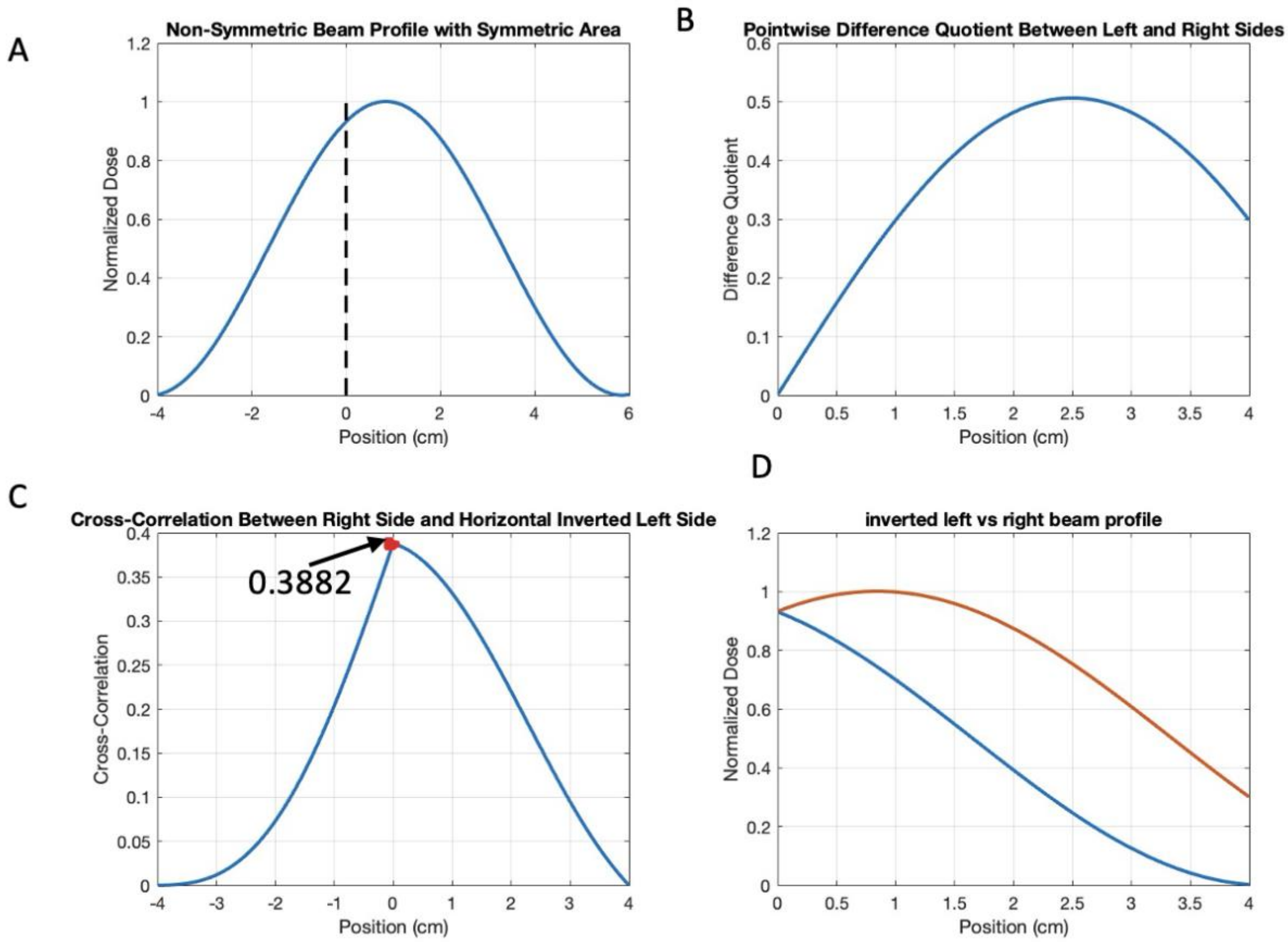

**Figure 4: A)** Simulation plot of a non-symmetric linac beam by superimposing various sinusoids of different amplitudes and phases**)** Calculation of Pointwise Difference Quotient (PDQ) between left and right sides of profile with respect to position from center of the beam **C)** Cross-correlation (CSI) values between vertically flipped left and right sides of the profile with respect to lag in terms of the profile position, which is the amount of shift of the mirror left profile before correlating the two sides of the beam profile **D)** Comparison between the profiles of the right and vertically flipped left side of the simulated beam profile.

The intensity values of the beam profile were normalized and shifted, such that the minimum value was zero (**Figure 4a**). The corresponding area-based symmetric value was calculated to be 0.3306. The corresponding Pointwise Difference Quotient between the left and right for a position value of anything within the central 80% of the field was greater than 0.40 (**Figure 4b)**. The corresponding correlative symmetry index was 0.3882 for a lag of zero between the vertically flipped left profile and the right profile (**Figure 4c**). The resulting SSIM (Structural Similarity index measurement) between the left and right profiles was 0.6449. The most valuable correlative symmetric index was located at a position-based shift of zero. This is where both the inverted left and right profiles start at the initial position of the center of the entire beam profile. The other shift location positions start values for the inverted left and right profiles that no longer match the center of the whole beam profile. The value of the correlation that corresponds to a position shift of zero can be automatically selected to find the correlation symmetric index that corresponds to the similarity of the right and inverted left profiles.

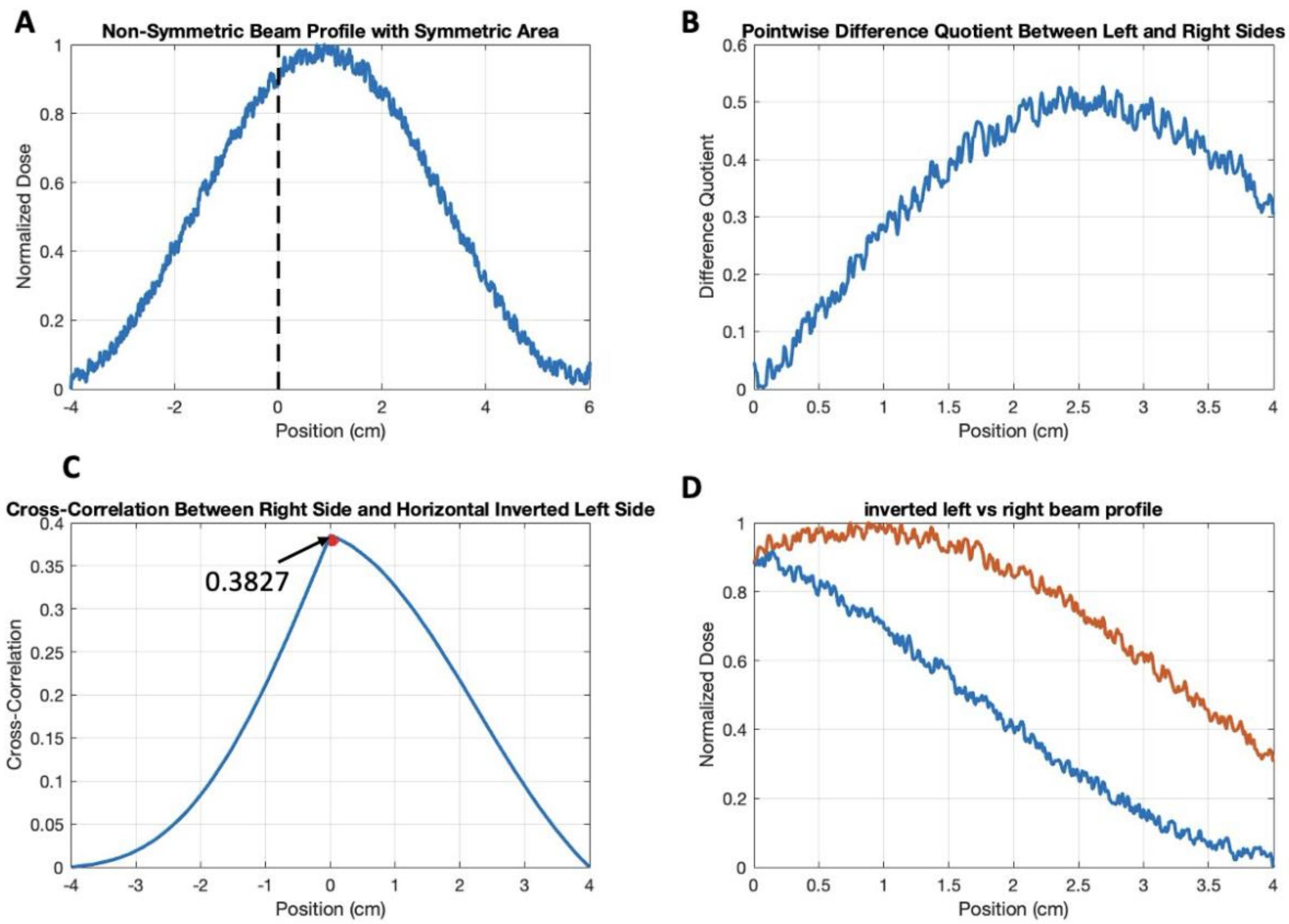

**Figure 5: A)** Simulation of a non-symmetric linac beam with roughly symmetric area with low passed filtered Gaussian noise superimposed **B)** Calculation of Pointwise Difference Quotient between left and right sides of profile with respect to position with Gaussian noise **C)** Cross-correlation values between vertically flipped left and right sides of the profile with Gaussian noise with respect to lag in terms of the profile position, which is the amount of shift of the mirror left profile before correlating the two sides of the beam profile **D)** Comparison between the profiles of the right and vertically flipped left side of the simulated beam profile with Gaussian noise.

To examine the impact of noisy data, low-pass filtered Gaussian noise was added to the beam profile to simulate the electronic noise from the ion chamber and associated electronics **(Figure**

**5A)**. The calculated area-based symmetric value under noisy conditions is 0.312. The corresponding Pointwise Difference Quotient between the left and right for a position value of anything within the central 80% of the field was greater than 0.45 **(Figure 5B)**. The correlative index between the left and right profiles was 0.3827 **(Figure 5C)**. The resulting SSIM (Structural Similarity index measurement) between the left and right profiles was in 0.5401.

The average percentage change for each of the pointwise difference quotients , area-based symmetric values, SSIM, and correlative index were 0.111, 0.056, 0.1625, and 0.0143, respectively, for different modelled noises. The pairwise t-statistic between the different obtained percentage changes for each standard metric compared to the proposed CSI metric under different modelled noises showed significant differences, as shown in Figure 6. The correlative index metric was the least impacted by, noise while the pointwise difference quotient and SSIM were the most impacted by noise. The left and right beam profiles are compared in **Figure 5D**.

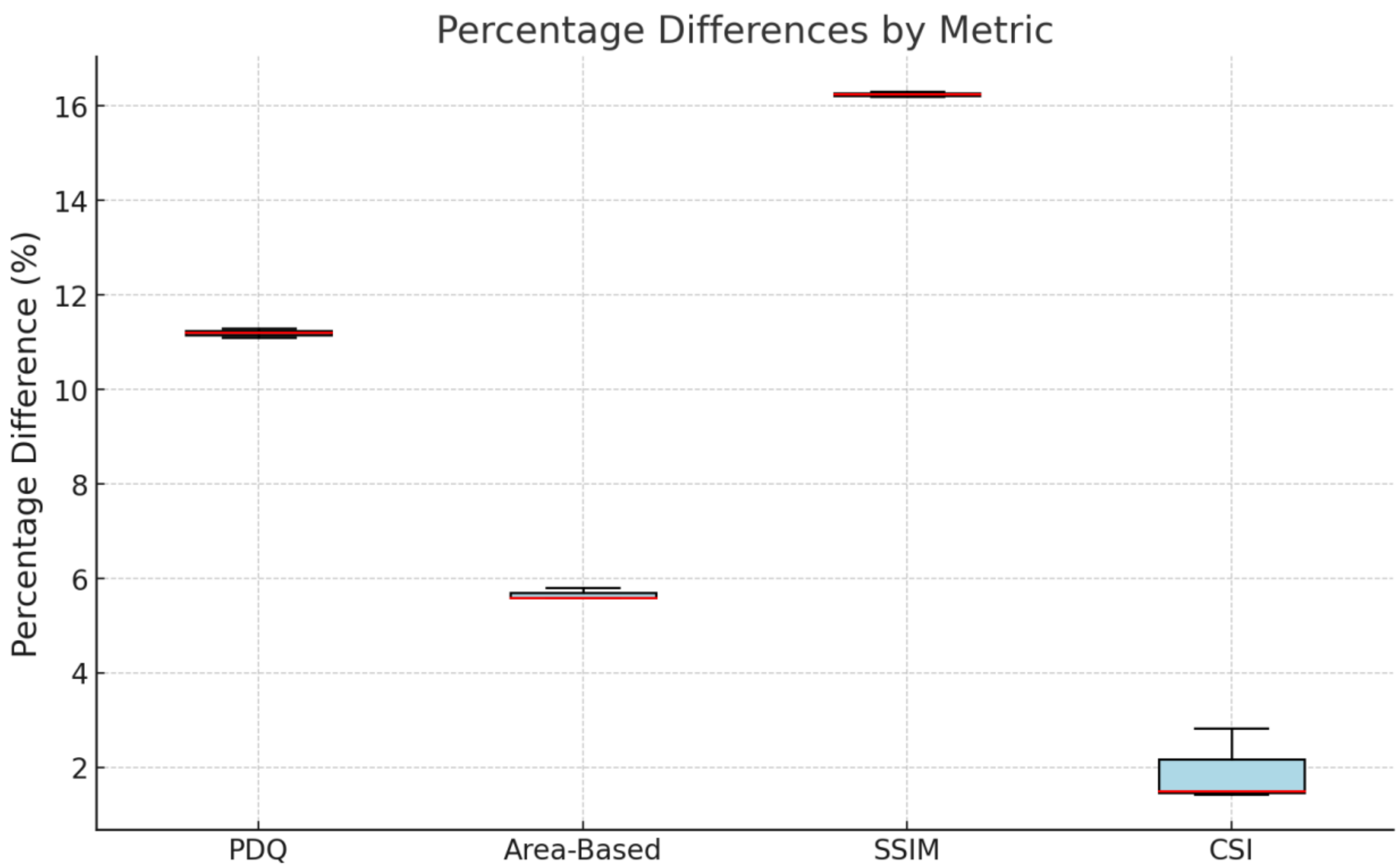

| Comparison | t-statistic | p-value | Significant (p < 0.05) |
|---|---|---|---|
| CSI vs PDQ | -20.0621 | 0.0021 | TRUE |
| CSI vs Area | -8.0745 | 0.0133 | TRUE |
| CSI vs SSIM | -31.1662 | 0.001 | TRUE |

**Figure 6:** Demonstrates that CSI had statistically significant difference in percentage change between non-symmetry noise and non-symmetric no-noise profiles compared to the other metrics.

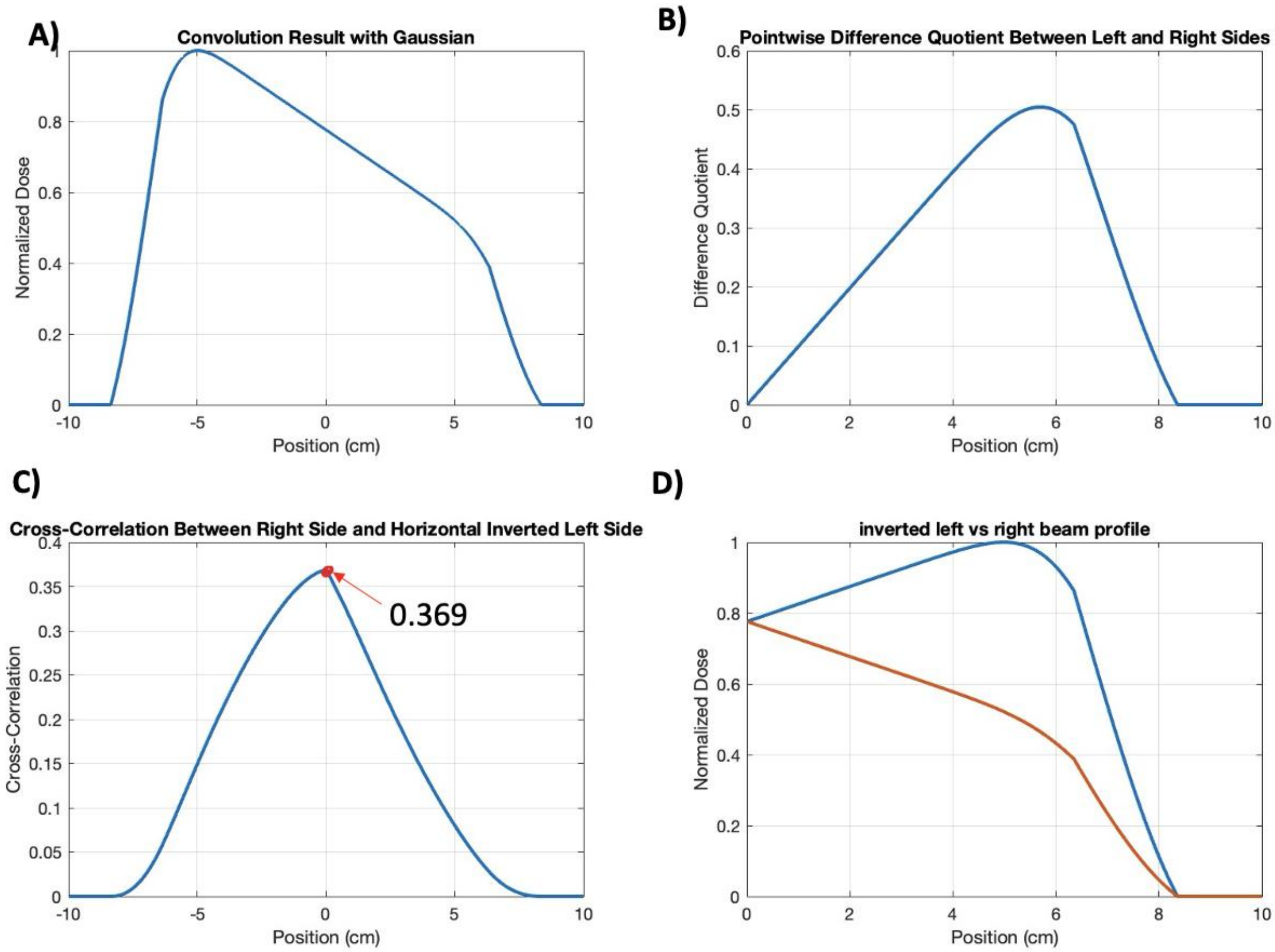

**Figure 7: A)** Simulation of a non-symmetric linac beam based on Wedge effects **B)** Calculation of Pointwise Difference Quotient between left and right sides of profile with respect to position **C)** Cross-correlation (CSI) values between vertically flipped left and right sides of the profile with respect to lag in terms of the profile position, which is the amount of shift of the mirror left profile before correlating the two sides of the beam profile **D)** Comparison between the profiles of the right and vertically flipped left side of the simulated beam profile with Wedge Effects.

The calculated area-based symmetric value was 0.225. The corresponding Pointwise Difference Quotient between the left and right for a position value of anything within the central 80% of the field was greater than 0.30 **(Figure 7B)**. The correlative index between the left and right profiles was 0.369 **(Figure 7C)**. The resulting SSIM (Structural Similarity index measurement) between the left and right profiles was 0.8814.

| Non-symmetric (Area-based metric) | Non-symmetric (Point-wise difference quotient) | Non-symmetric (Correlative Symmetric Index) | Non-symmetric (SSIM) | Non-symmetric Metric from Previous Study |
|---|---|---|---|---|
| **0.225** | >= 0.30 | 0.369 | 0.8814 | 0.4 [8] |

**Table 1:** Comparison of calculated symmetric based methods for a non-symmetric beam profile of a wedge of 15 degrees with previous published literature non-symmetric beam profile.

***Results obtained from Measured Beam PDD profiles.***

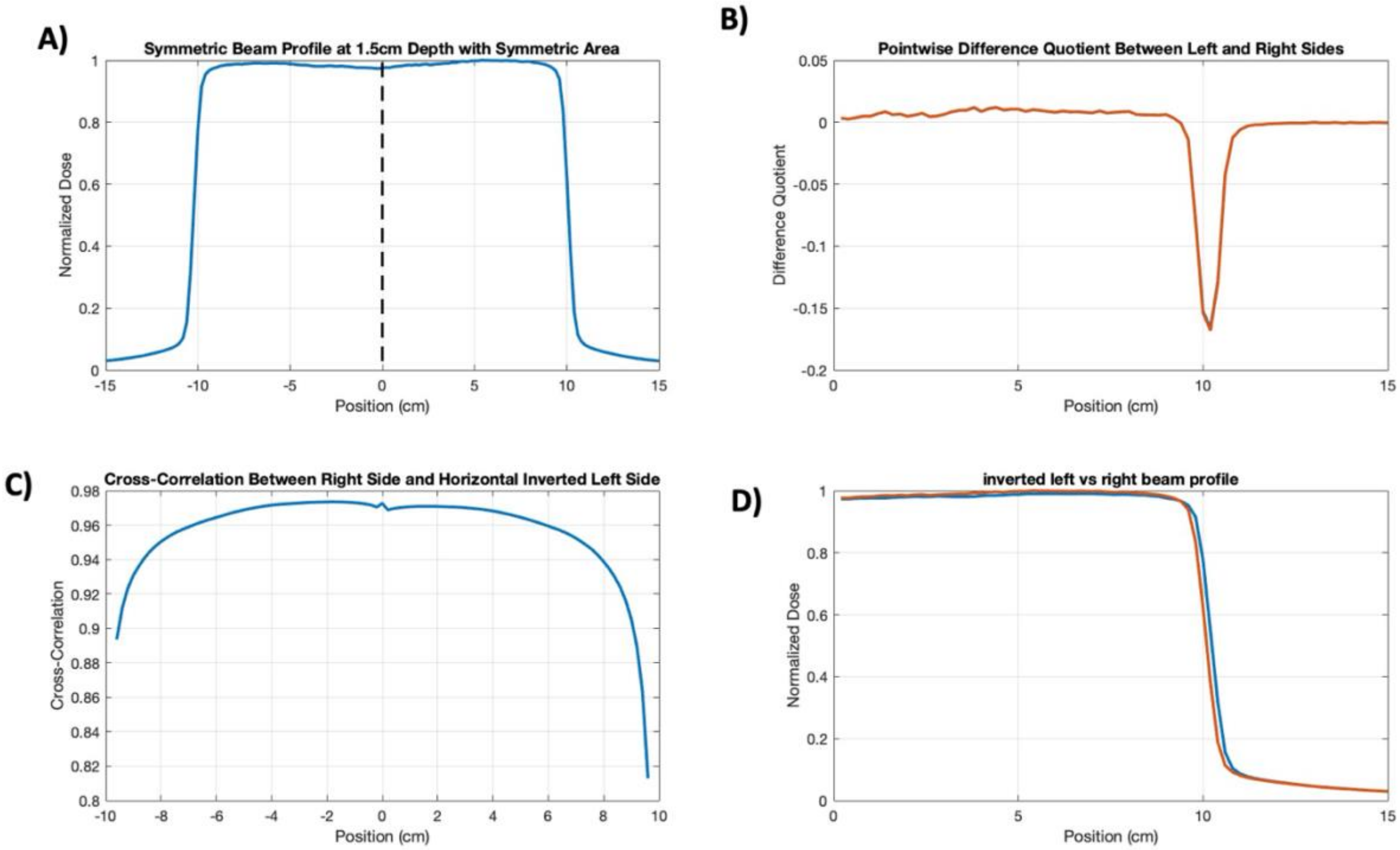

**Figure 8: A)** Measured symmetric PPD beam at 1.5cm Depth with roughly symmetric area **B)** Calculation of pointwise difference quotient between left and right sides of profile with respect to position **C)** Cross-correlation (CSI) values between vertically flipped left and right sides of the profile with respect to lag in terms of the profile position, which is the amount of shift before correlating the two sides of the beam profile **D)** Comparison between the profiles of the right and vertically flipped left side of the beam profile.

The beam at 1.5 cm depth **(Figure 8A)** had an area-based symmetric value of 0.0028, while the Pointwise Difference Quotient between the left and right was 0.150 within the central 80% of the field **(Figure 8B)**. The Correlation Symmetry Index was 0.974, indicating a highly symmetric beam profile **(Figure 8C)**. The left and right beam profiles were compared visually for qualitative assessment of symmetry, as shown in **Figure 8D**. The resulting SSIM value between the right and left profiles was 0.9841.

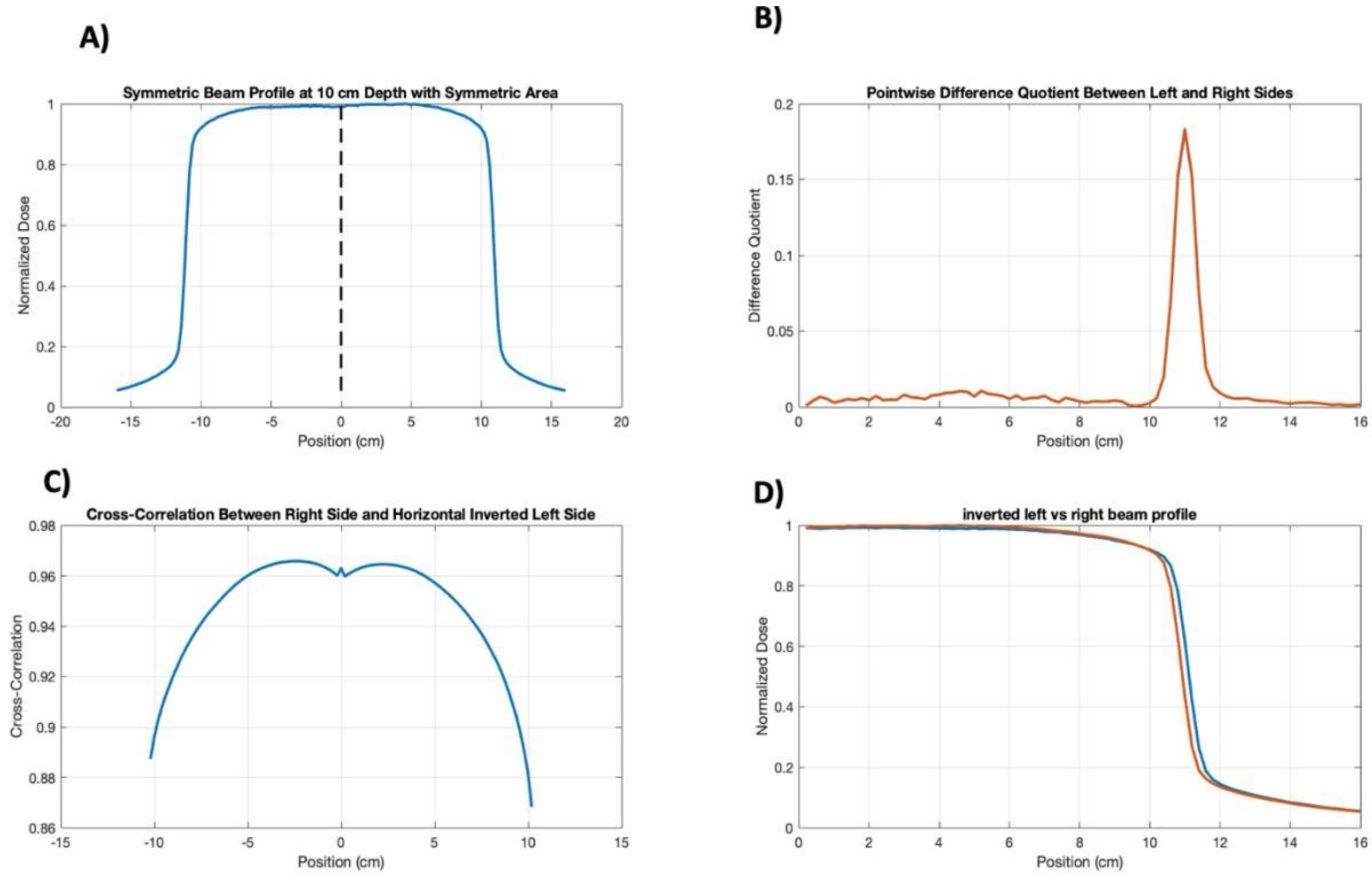

**Figure 9: A)** Measured symmetric PPD beam at 10 cm Depth with roughly symmetric area **B)** Calculation of pointwise difference quotient between left and right sides of profile with respect to position **C)** Cross-correlation (CSI) values between vertically flipped left and right sides of the profile with respect to lag in terms of the profile position, which is the amount of shift before correlating the two sides of the beam profile **D)** Comparison between the profiles of the right and vertically flipped left side of the beam profile.

The beam profile at a depth of 10 cm **(Figure 9A)** had an area-based symmetric value of 0.0046 with a Pointwise Difference Quotient of 0.18 **(Figure 9B)** within the central 80% of the field. CSI was evaluated to be 0.964, again very close to symmetric **(Figure 9C)**. The left and right beam profiles are compared in **Figure 9D**. Resulting SSIM values for the left and right beam profileswas 0.9815 .

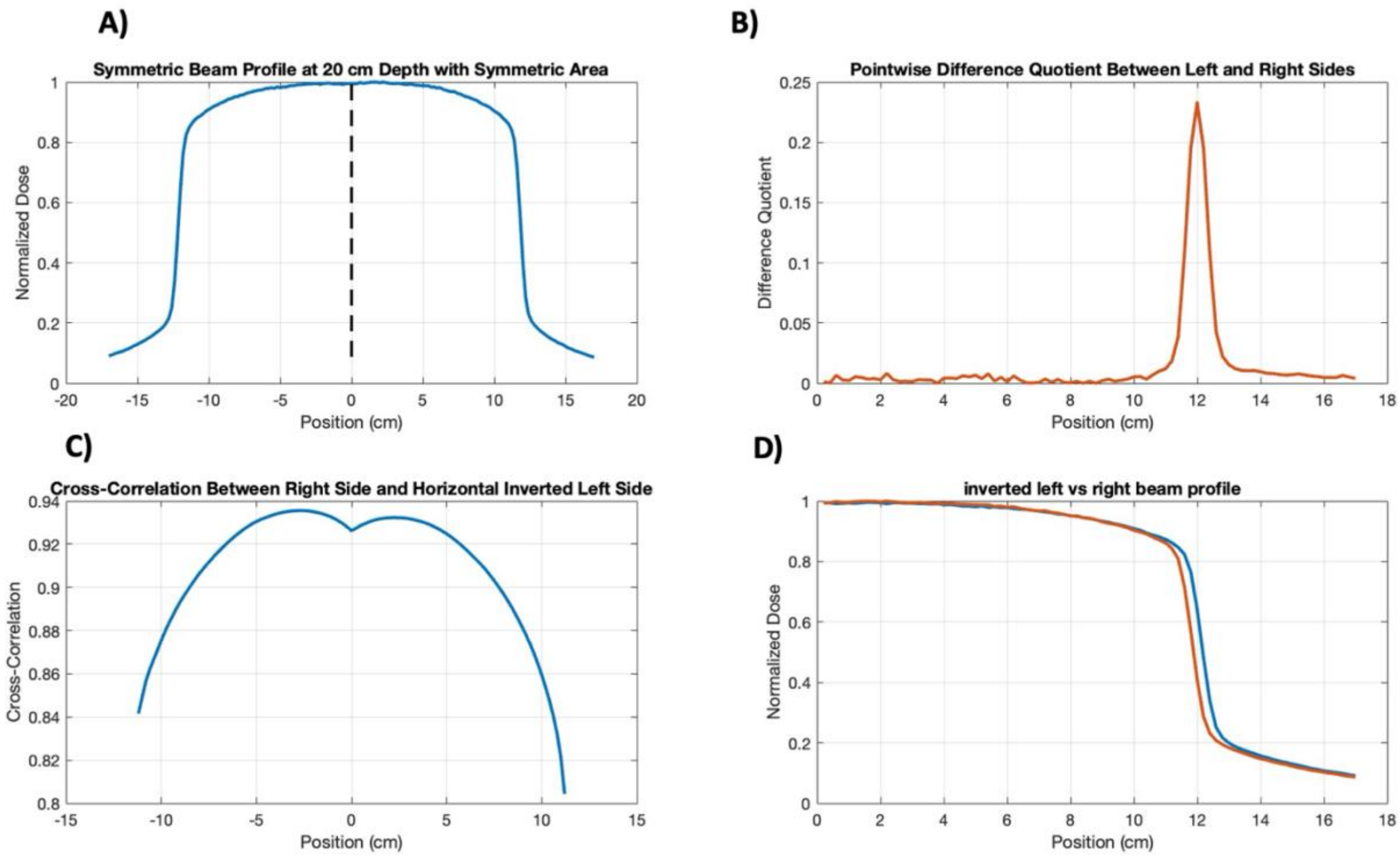

**Figure 10: A)** Measured symmetric PPD beam at 20 cm Depth with roughly symmetric area **B)** Calculation of pointwise difference quotient between left and right sides of profile with respect to position **C)** Cross-correlation (CSI) values between vertically flipped left and right sides of the profile with respect to lag in terms of the profile position, which is the amount of shift before correlating the two sides of the beam profile **D)** Comparison between the profiles of the right and vertically flipped left side of the beam profile.

The beam at a depth of 20 cm **(Figure 10A)** had an area-based symmetric value of 0. The Pointwise Difference Quotient between the left and right profiles is 0.23 within the central 80% of the field. **(Figure 10B)**. The correlation index was 0.93, which is very close to 1, as expected **(Figure 10C)** because this profile is close to symmetric except for the slight difference in penumbras of the left and right profiles, as shown in **Figure 10D**. Resulting SSIM values for the left and right beam profiles was 0.9690.

***Relationship Between Clinical Tolerances for PDQ, Area-Based Symmetry, and CSI***

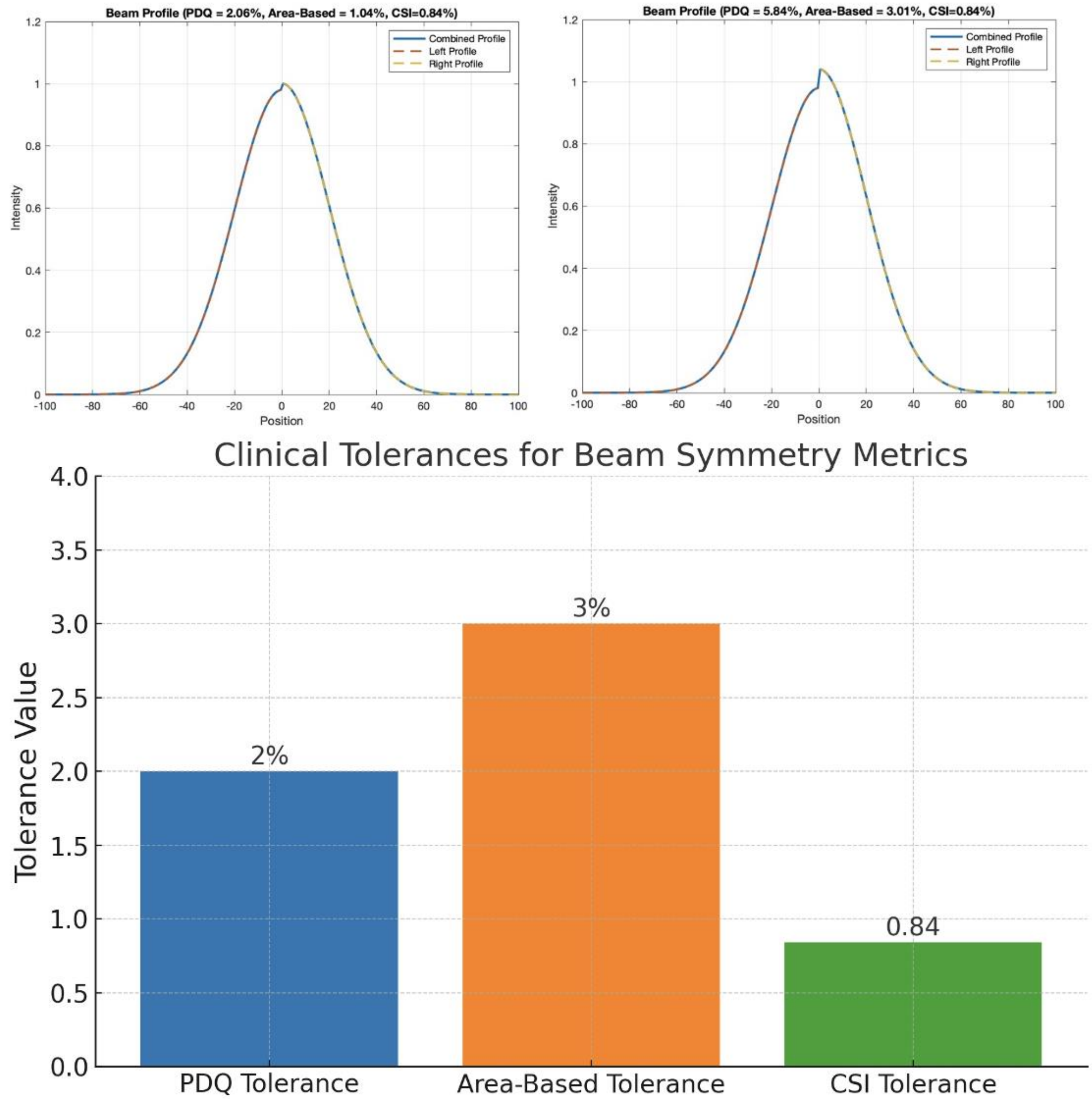

**Figure 11:** Shows the corresponding CSI clinical tolerance (0.84) in relationship to standard clinical tolerances for modelled beam profiles with PDQ (±2%)[9] and area-based (±3%) metrics[10]. This shows that for a profile to be clinically acceptable in symmetry when measuring symmetry with proposed CSI one need to have a CSI value above 0.84.

The clinical tolerances for beam symmetry are typically defined based on the PDQ (±2%) and area-based metrics (±3%) to ensure that the delivered dose profile aligns closely with the planned profile. These metrics capture different aspects of beam symmetry; the PDQ is sensitive to point-wise deviations, whereas area-based metrics reflect the integrated differences across the entire profile. However, these traditional metrics do not fully account for the spatial distribution and correlations within the beam profile, potentially overlooking subtle, but clinically significant asymmetries. In contrast, CSI incorporates both local and global structural information, making it more robust to high-frequency noise and more reflective of true beam symmetry.

From the model beam profiles provided, it is evident that a beam profile with a PDQ of exactly 2% and an area-based metric of exactly 3% corresponds to a CSI of approximately 0.84. This suggests that a CSI tolerance of approximately 0.84 may be a suitable threshold for clinical quality assurance, aligning well with the established tolerances for PDQ and area-based methods. This value represents a practical threshold where the CSI captures not only the local deviations emphasized by the PDQ but also the integrated symmetry emphasized by area-based approaches, potentially providing a more comprehensive measure of beam quality. This alignment can facilitate the clinical adoption of CSI as a complementary or alternative metric for beam symmetry assessment in RT (radiation therapy).

## Discussion

Standard methods for quantifying the beam symmetry have major limitations. The pointwise difference quotient and area-based methods were significantly affected by the amount of noise in the beam profile. In terms of symmetry through the area-based method, there is no necessary guarantee because the left profile area equals the right profile area, where the beam profile is symmetric. The proposed technique for finding the correlation index of the left and right profiles of the beam is less affected owing to noise levels than the area since it is a direct quantification of the similarity of the left and right profiles of the beam. The CSI bias achieved comparable results in terms of quantifying symmetry for high asymmetric profiles to the results obtained through the PDQ and area-based symmetric metrics for the modelled non-symmetric beam profile scenarios. asFor the realistic non-symmetric beam model scenario, the calculated area-based symmetric value was 0.225, and the pointwise difference quotient between the left and right beam profiles for anything between the central and the half maximum point was equal to or less than 0.500. The corresponding CSI-biased metric for the realistic non-symmetric beam modelled was 0.369, which lies between 0.225 and 0.500. These results are comparable to the asymmetric values obtained from distributions with a wedge transmission factor[8] (Table 1).

One of the most important aspects of safe and effective radiotherapy is the accurate quantification of the radiation beams. Therefore, quality assurance is heavily emphasized in the field. Properly evaluating beam symmetry is crucial in clinical settings for delivering radiotherapy treatments, ensuring that the prescribed dose covers the entire tumor uniformly without overdelivering to any peripheral tissues. Therefore, it is necessary to utilize effective mathematical measures that are representative of the beam symmetry.

## Clinical Relevance and Interpretability of the Correlative Symmetric Index (CSI)

The Correlative Symmetric Index (CSI) offers a complementary approach to conventional metrics, such as the Point Dose Quotient (PDQ) and area-based symmetry metrics, by capturing both pointwise and area-integrated differences while being inherently more robust to high-frequency noise. This is particularly valuable in clinical contexts, where beam profiles exhibit subtle but clinically significant asymmetries that might be masked by noise or minor setup errors. Unlike PDQ, which captures only localized point-to-point differences, and area-based methods that integrate absolute differences over the beam cross-section, CSI can incorporate spatial correlations across the entire profile, providing a more holistic assessment of beam symmetry. This is crucial in applications such as stereotactic radiosurgery (SRS) or intensity-modulated radiotherapy (IMRT), where precise dose distributions are critical to avoid underdosing or overdosing of nearby organs at risk (OARs) (Table 2). Furthermore, CSI can be more sensitive to changes in the beam profile shape that reflect equipment miscalibration, gantry sag, or multileaf collimator (MLC) errors, offering an early indicator of potential machine performance issues. To further bridge the gap between mathematical robustness and clinical applicability, CSI can be adapted to reflect the same clinical tolerances as PDQ and area-based metrics, such as the 2% PDQ and 3% area-based tolerances used in standard quality assurance protocols. This alignment allows for easier clinical adoption and interpretation, making it a potentially valuable addition to the Beam QA Toolkit.

| Metric | Strengths | Weaknesses | Clinical Relevance |
|---|---|---|---|
| PDQ | Simple, intuitive, widely adopted | Susceptible to noise, only captures point-to-point deviations | Effective for gross symmetry errors in large fields |
| Area-Based Symmetry | Captures integrated differences, less sensitive to single-point noise | Can miss localized asymmetries, sensitive to profile width variations | Useful for general beam shape validation |
| CSI (Proposed) | Robust to noise, captures spatial correlations, sensitive to profile shape | More complex to calculate, requires clear clinical tolerance definitions | Ideal for high-precision treatments like SRS, IMRT, and MR-Linac adaptive workflows |

**Table 2:** Description showing the comparison of strength, weaknesses and clinical relevance of the standard methods vs the CSI proposed metric.

### *Limitations*

While the Correlative Symmetric Index (CSI) demonstrates improved robustness to noise and greater sensitivity to profile asymmetry compared to PDQ and area-based metrics, several limitations should be acknowledged. First, the CSI threshold of 0.84, proposed as a clinical tolerance equivalent to ±2% PDQ and ±3% area-based symmetry, is empirically derived from modelled profiles and requires broader validation across various beam energies, modalities, and detector configurations. Second, the CSI computation involves cross-correlation analysis, which is more computationally intensive than traditional point-wise comparisons, potentially limiting real-time deployment without appropriate algorithmic optimization. Third, CSI, similar to all profile metrics, assumes accurate spatial registration between mirrored halves of the beam profile; any misalignment or geometric distortion may influence the results. Lastly, while CSI offers quantitative advantages, its clinical integration would benefit from prospective multi-institutional studies to assess inter-device reproducibility and its performance in real-world QA workflows. Future work should explore adaptive thresholding, extension to 2D or volumetric beam profiles, and implementation of clinical software.

## Conclusion

In this study, standard quantitative mathematical models for beam symmetry were assessed and compared with the proposed correlative index for both simulated and measured beam profiles. The proposed correlative index proved to be more resistant to noise than standard area and pointwise methods as a measure of beam symmetry. More representative beam symmetry metrics are required for assess beam symmetry in quality assurance, as minor errors in representation symmetry can affect treatment outcomes. CSI can be integrated into routine QA software to enhance beam symmetry verification under noisy conditions.